\documentclass[10pt,a4paper]{article}

\usepackage[utf8]{inputenc}
\usepackage[T1]{fontenc}
\usepackage{amsmath,amssymb}
\usepackage{graphicx}
\usepackage{booktabs}
\usepackage{hyperref}
\usepackage{xcolor}
\usepackage{listings}
\usepackage[margin=1in]{geometry}
\usepackage{natbib}
\usepackage{multirow}
\usepackage{array}
\usepackage{pgfplots}
\pgfplotsset{compat=1.18}

\hypersetup{
    colorlinks=true,
    linkcolor=blue,
    citecolor=blue,
    urlcolor=blue
}

\title{VIDS: A Verified Imaging Dataset Standard for Medical AI}

\author{
    Dr.~Joan S.~Muthu, PhD\thanks{Corresponding author: standards@vidsstandard.org} \\
    John Shalen \\
    Princeton Medical Systems \\
    \texttt{standards@vidsstandard.org}
}

\date{}

\begin{document}

\maketitle

\begin{abstract}
Medical imaging AI development is fundamentally dependent on annotated datasets,
yet no existing standard provides machine enforceable validation across
dataset structure, annotation provenance, quality documentation, and ML
readiness within a single framework.
DICOM standardizes image acquisition, storage, and communication at the
individual study level. BIDS organizes neuroimaging research datasets
with consistent naming conventions. Neither of these standards addresses the
curation layer, viz., who annotated what, when, with what tool, and to what
quality standard.

This paper presents VIDS (Verified Imaging Dataset Standard), an open
specification that defines folder layout, file naming, annotation
provenance schemas, quality documentation, and 21 machine enforceable
validation rules across two compliance profiles. VIDS uses NIfTI as a
canonical working format while preserving full DICOM metadata in sidecars
for traceability, and supports export to any downstream ML framework
(nnU-Net, MONAI, COCO, flat NIfTI) without loss of provenance.

Twenty two compliance dimensions are defined and four major public
datasets, viz., LIDC-IDRI, BraTS, CheXpert, and the Medical Segmentation
Decathlon, are benchmarked against these dimensions. It is observed that
even widely used datasets satisfy only 20--39\% of these dimensions, with
provenance and quality documentation as the largest systematic gaps.
Further, LIDC-Hybrid-100 is released as a 100 subject VIDS compliant
reference CT dataset with consensus segmentation masks from four radiologist
annotations (mean pairwise Dice 0.7765), which validates 21/21 on the Full
compliance profile.

VIDS is fully open source: the specification is CC~BY~4.0, all tools are
Apache~2.0, the reference validator is available on PyPI
(\texttt{pip install vids-validator}), and LIDC-Hybrid-100 is published
on Zenodo (\url{https://doi.org/10.5281/zenodo.19582717}).
\end{abstract}

\section{Introduction}
\label{sec:introduction}

The medical imaging AI pipeline is dependent on annotated datasets at every
stage, from algorithm development through regulatory submission. Nevertheless,
there is no standard that governs how these datasets should be structured,
documented, and validated. It is a common observation that teams receive
datasets as opaque archives of unnamed NIfTI files with undocumented
annotations, and consequently spend days or weeks on data engineering before
any model training can commence.

Existing standards solve adjacent problems at different layers.
DICOM~\citep{dicom} standardizes image acquisition, storage, and
communication. BIDS~\citep{gorgolewski2016} organizes neuroimaging datasets
with consistent naming conventions and sidecar JSON metadata.
COCO~\citep{lin2014coco} and Pascal VOC~\citep{everingham2010pascal}
define annotation formats for natural image detection tasks.
Datasheets for Datasets~\citep{gebru2021datasheets} and Data
Cards~\citep{pushkarna2022data} propose documentation frameworks for
dataset transparency. These frameworks provide optional free text sections
for describing annotation processes but do not enforce structured
per annotation provenance or validate completeness automatically.
Hence, none of these standards provide machine enforceable validation across
the full scope of an annotated medical imaging dataset, viz., directory
structure, file naming, annotation provenance (who annotated, when, with
what tool, under what quality controls), quality documentation
(inter annotator agreement, class distributions), and ML readiness
(train/val/test splits with leakage prevention).

VIDS fills this gap. It defines a complete directory structure for
multi modality imaging datasets, mandatory per annotation provenance
documentation, 21 machine enforceable validation rules, two compliance
profiles for incremental adoption, and format agnostic export to
downstream ML frameworks. It is noteworthy that VIDS is a \textit{delivery
structure} and not a data product. It enforces that quality is documented,
not that quality is good. A bad dataset in VIDS format is still bad, but
\textit{visibly} bad: thin provenance, low inter annotator agreement, and
demographic gaps are exposed in structured, machine readable fields.
Therefore, VIDS shifts dataset evaluation from implicit trust to explicit,
inspectable evidence.

The contributions of this paper are as follows:
\begin{itemize}
    \item[(i)] The VIDS v1.0 specification, covering dataset structure
    through ML readiness (Section~\ref{sec:design}).
    \item[(ii)] A zero dependency reference validator published on
    PyPI (Section~\ref{sec:implementation}).
    \item[(iii)] A compliance analysis benchmarking four major public
    datasets against 22 VIDS dimensions (Section~\ref{sec:analysis}).
    \item[(iv)] LIDC-Hybrid-100, a public VIDS compliant reference
    dataset with 100 subjects, 89 annotated, 224 nodules, and full
    provenance (Section~\ref{sec:case_study}).
\end{itemize}

\section{Related Work}
\label{sec:related}

\subsection{Medical Imaging Standards}

The DICOM standard~\citep{dicom} is the foundation of medical image
acquisition, storage, and communication, defining how scanners produce
and transmit images. DICOM operates at the study level and does not
address how annotated datasets should be organized for AI development.
DICOM SEG and DICOM SR extend DICOM for segmentation storage and
structured reporting, but template variability across vendors limits
their utility as a universal dataset standard.

The Brain Imaging Data Structure (BIDS)~\citep{gorgolewski2016}
introduced consistent naming conventions (\texttt{sub-},
\texttt{ses-}) and sidecar JSON metadata for organizing neuroimaging
datasets. VIDS adopts these conventions and extends them to
multi modality annotation with mandatory provenance. Nevertheless, BIDS
does not cover annotation provenance, quality documentation, or automated
compliance validation, and is specific to neuroimaging.

\subsection{Annotation Formats and Documentation}

COCO~\citep{lin2014coco} and Pascal VOC~\citep{everingham2010pascal}
define annotation schemas for object detection in natural images. These
formats handle bounding boxes and segmentation polygons but lack
provenance metadata and are not designed for volumetric medical imaging.

Datasheets for Datasets~\citep{gebru2021datasheets} and Data
Cards~\citep{pushkarna2022data} propose documentation frameworks for
dataset transparency. More recently, RSNA launched ATLAS (2025), a
searchable catalog of standardized dataset and model cards for medical
imaging AI. These frameworks serve the purpose of documenting datasets,
describing annotation processes, limitations, and intended uses. However,
they do not define or enforce dataset file structure. A dataset card with
thorough free text documentation can accompany a completely unstructured
archive. VIDS operates at a different layer: it defines the structure
itself and validates compliance automatically.

\subsection{The Gap}

Table~\ref{tab:comparison} summarizes the coverage of existing standards.
It is evident that no existing standard combines all five requirements for
production medical imaging AI datasets: directory structure, annotation
provenance, quality documentation, ML readiness, and automated validation.

\begin{table}[h]
\centering
\caption{Coverage comparison across five dataset requirements.}
\label{tab:comparison}
\small
\begin{tabular}{lccccc}
\toprule
\textbf{Standard} & \textbf{Structure} & \textbf{Provenance} & \textbf{Quality} & \textbf{ML Ready} & \textbf{Validator} \\
\midrule
DICOM         & Partial & ---     & ---     & ---     & ---     \\
BIDS          & Yes     & ---     & ---     & ---     & Yes     \\
COCO / VOC    & Partial & ---     & ---     & ---     & ---     \\
Datasheets / Cards & ---  & Partial & Partial & ---   & ---     \\
RSNA ATLAS    & ---     & Partial & ---     & ---     & ---     \\
\textbf{VIDS} & \textbf{Yes} & \textbf{Yes} & \textbf{Yes} & \textbf{Yes} & \textbf{Yes} \\
\bottomrule
\end{tabular}
\end{table}

\section{VIDS Design}
\label{sec:design}

\subsection{Design Principles}

VIDS is built on five principles:

\begin{itemize}
    \item[(i)] \textbf{Provenance is mandatory, not optional.} Every
    annotation must document who created it, when, with what tool, and
    under what quality controls.
    \item[(ii)] \textbf{Validation is automated.} Compliance is
    determined by running a validator, not by reading a checklist. If the
    validator passes, the dataset is compliant.
    \item[(iii)] \textbf{The standard is format agnostic at delivery.}
    VIDS defines a canonical internal structure. Datasets can be exported
    to any downstream framework without loss of provenance.
    \item[(iv)] \textbf{Profiles enable incremental adoption.} The POC
    profile requires 15 rules for quick prototypes. The Full profile
    requires all 21 for production, publications, and regulatory
    submissions.
    \item[(v)] \textbf{The standard complements, not replaces.} VIDS
    works with NIfTI (file format), DICOM (source data), BIDS (naming
    conventions), and existing ML frameworks.
\end{itemize}

\subsection{Dataset Structure}

A VIDS dataset organizes data into a root containing metadata files, a
\texttt{sub-*/ses-*/modality/} hierarchy for source imaging, a
\texttt{derivatives/annotations/} tree mirroring the source layout, and
\texttt{quality/} and \texttt{ml/} directories for Full profile
compliance. File naming follows the pattern:

\begin{center}
\texttt{sub-<ID>\_ses-<ID>\_<modality>\_<suffix>.<ext>}
\end{center}

The complete directory structure is shown in Figure~\ref{fig:structure}.

\begin{figure}[h]
\centering
\begin{lstlisting}[basicstyle=\ttfamily\footnotesize,frame=none,backgroundcolor=\color{white}]
dataset/
  .vids                         # Profile marker
  dataset_description.json      # 6 required fields
  participants.json             # Subject demographics
  README.md
  sub-001/ses-baseline/ct/
    sub-001_ses-baseline_ct_img.nii.gz
    sub-001_ses-baseline_ct_img.json
  derivatives/annotations/sub-001/ses-baseline/ct/
    sub-001_ses-baseline_ct_seg.nii.gz
    sub-001_ses-baseline_ct_seg.json
  quality/                      # Full profile
    quality_summary.json
    annotation_agreement.json
  ml/                           # Full profile
    splits.json
\end{lstlisting}
\caption{VIDS directory structure. Root level metadata files, subject/session/modality hierarchy, annotations tree, quality documentation, and ML splits.}
\label{fig:structure}
\end{figure}

\subsection{Annotation Provenance}

The annotation sidecar JSON (\texttt{*\_seg.json}) is the core of VIDS
provenance tracking. It contains a \texttt{Provenance} object with three
sub objects:

\begin{itemize}
    \item \textbf{Annotator:} identity (\texttt{ID} or \texttt{Name}),
    credentials, specialty, institution
    \item \textbf{AnnotationProcess:} tool, version, date, time spent,
    method (manual, semi automated, automated)
    \item \textbf{QualityControl:} reviewer identity, review date,
    outcome (approved/revisions/rejected), confidence score
\end{itemize}

At minimum, each annotation must document annotator identity (at least
one of \texttt{ID} or \texttt{Name}) and either annotation date or tool.
This threshold is intentionally low. The goal is to make provenance
\textit{assessable}, not to mandate any particular level of documentation.
A sidecar with only an annotator ID and a tool name passes validation,
but anyone reviewing it can immediately perceive the gap and request
further detail.

Figure~\ref{fig:provenance} shows a representative annotation sidecar.

\begin{figure}[h]
\centering
\begin{lstlisting}[basicstyle=\ttfamily\footnotesize,language={}]
{
  "VIDSVersion": "1.0",
  "AnnotationType": "segmentation",
  "SourceImage": "sub-001_ses-baseline_ct_img.nii.gz",
  "LabelMap": {"0": "background", "1": "nodule"},
  "Provenance": {
    "Annotator": {
      "ID": "radiologist_001",
      "Credentials": "MD, Board-certified, 8yr"
    },
    "AnnotationProcess": {
      "Tool": "3D Slicer 5.6.2",
      "Date": "2026-03-15",
      "TimeSpent_minutes": 18
    },
    "QualityControl": {
      "ReviewedBy": "senior_radiologist_001",
      "ReviewOutcome": "approved",
      "Confidence": 0.93
    }
  }
}
\end{lstlisting}
\caption{VIDS annotation sidecar with provenance, documenting the
annotator, tool, date, and QC review for each annotation.}
\label{fig:provenance}
\end{figure}

\subsection{Validation Rules}

VIDS defines 21 validation rules across six categories
(Table~\ref{tab:rules}). Each rule produces one of four outcomes:
\textbf{PASS} (satisfied), \textbf{FAIL} (violated, dataset is
non compliant), \textbf{WARN} (recommended practice not followed), or
\textbf{SKIP} (not applicable to the declared profile). A dataset is
VIDS compliant if and only if zero rules have FAIL status.

\begin{table}[h]
\centering
\caption{VIDS 21 validation rules. POC profile enforces S, I, A, and D
rules (15 checks). Full profile enforces all 21.}
\label{tab:rules}
\small
\begin{tabular}{llll}
\toprule
\textbf{Rule} & \textbf{Category} & \textbf{Check} & \textbf{Profile} \\
\midrule
S001 & Structure   & \texttt{.vids} marker exists                          & POC + Full \\
S002 & Structure   & \texttt{dataset\_description.json} valid (6 fields)   & POC + Full \\
S003 & Structure   & \texttt{participants.json} or \texttt{.tsv} exists    & POC + Full \\
S004 & Structure   & \texttt{README.md} exists                             & POC + Full \\
S005 & Structure   & Subject directories (\texttt{sub-*}) exist            & POC + Full \\
S006 & Structure   & Session directories (\texttt{ses-*}) exist            & POC + Full \\
\midrule
I001 & Imaging     & NIfTI files present per subject                       & POC + Full \\
I002 & Imaging     & Imaging sidecar JSONs present                         & POC + Full \\
I003 & Imaging     & Imaging sidecar JSONs are valid                       & POC + Full \\
I004 & Imaging     & VIDS naming convention (WARN)                         & POC + Full \\
\midrule
A001 & Annotation  & \texttt{derivatives/annotations/} exists              & POC + Full \\
A002 & Annotation  & Segmentation files exist                              & POC + Full \\
A003 & Annotation  & Annotation sidecar JSONs exist                        & POC + Full \\
A004 & Annotation  & Annotation JSONs valid + \texttt{VIDSVersion}         & POC + Full \\
A005 & Annotation  & Provenance fields populated                           & POC + Full \\
\midrule
Q001 & Quality     & \texttt{quality/} directory exists                    & Full only \\
Q002 & Quality     & \texttt{quality\_summary.json} present                & Full only \\
Q003 & Quality     & \texttt{annotation\_agreement.json} present           & Full only \\
\midrule
M001 & ML          & \texttt{ml/} directory exists                         & Full only \\
M002 & ML          & \texttt{ml/splits.json} present                       & Full only \\
\midrule
D001 & Metadata    & \texttt{CHANGES.md} exists (WARN)                     & POC + Full \\
\bottomrule
\end{tabular}
\end{table}

\subsection{Compliance Profiles}

The \textbf{POC profile} enforces 15 rules (S, I, A, D categories) for
quick prototypes, internal research, and pilot deliveries. The
\textbf{Full profile} enforces all 21 rules, adding quality
documentation (inter annotator agreement, quality summary) and ML
readiness (documented train/val/test splits). Organizations can start at
POC and graduate to Full as their processes mature. In production
settings, VIDS validation PASS can serve as an objective dataset
acceptance criterion in data procurement workflows.

\subsection{Extension Mechanism}

VIDS supports domain specific extensions without modifying the core
specification. Custom modality codes can be documented in
\texttt{dataset\_description.json}. The
\texttt{Annotations[].Characteristics} object is explicitly designed for
domain extension, and new fields can be added freely without affecting
validation. Custom derivative directories under
\texttt{derivatives/} are ignored by the validator.

\subsection{Export and Framework Compatibility}

VIDS achieves interoperability through framework compatibility, whereby
datasets curated in VIDS can be exported to any downstream ML framework
without restructuring the source or losing provenance.
Supported export targets include nnU-Net~\citep{isensee2021nnu},
MONAI~\citep{monai2020}, COCO JSON, and flat NIfTI layouts. All exports
include a \texttt{vids-provenance/} companion directory preserving the
full provenance chain. A mapping file links exported case identifiers
back to original VIDS subject IDs, ensuring traceability from model
input to annotation provenance.

\section{Reference Implementation}
\label{sec:implementation}

The VIDS reference validator (\texttt{validate\_vids.py}) is a
single file, zero dependency Python script compatible with Python~3.8+.
It reads the \texttt{.vids} marker to detect the compliance profile,
executes all applicable rules, and produces a structured JSON report
with per rule PASS/FAIL/WARN/SKIP status.

\begin{lstlisting}[language=bash]
# Install from PyPI
pip install vids-validator

# Validate a dataset
vids-validate /path/to/dataset --profile full
\end{lstlisting}

\begin{lstlisting}[language=Python]
# Programmatic use
from vids_validator import VIDSValidator
report = VIDSValidator("/path/to/dataset").validate()
assert report["Summary"]["Status"] == "PASS"
\end{lstlisting}

The validator integrates with CI pipelines via JSON output mode and
standard exit codes (0~=~pass, 1~=~fail). A browser based validator
supporting drag and drop zip validation is available at
\url{https://vidsstandard.org} for users without Python environments.
The validator executes in seconds for typical datasets (100+ subjects)
and introduces no dependencies beyond the Python standard library.

\section{Compliance Analysis}
\label{sec:analysis}

\subsection{Methodology}

Twenty two compliance dimensions are defined across six categories, and
four major public medical imaging datasets are assessed against these
dimensions. The dimensions correspond to the information that VIDS makes
structured and machine readable. Each dimension is scored as
\textit{satisfied} (present and machine readable), \textit{partial}
(present but not structured or requiring manual extraction), or
\textit{absent}. Partial scores count as 0.5 and are assigned only when
information is present in the dataset but not in a machine readable or
standardized form (e.g., provenance described in a companion paper but
absent from the dataset files). Scoring was performed against predefined
criteria for each dimension using publicly available dataset artifacts.
All scoring decisions are reproducible from the datasets' published
documentation and file structures.
Table~\ref{tab:dimensions} lists the 22 dimensions.

\begin{table}[h]
\centering
\caption{22 VIDS compliance dimensions across six categories.}
\label{tab:dimensions}
\small
\begin{tabular}{lp{5.5cm}}
\toprule
\textbf{Category} & \textbf{Dimensions} \\
\midrule
Structure (6) & Dataset marker, dataset description, participant registry, human readable README, subject hierarchy, session hierarchy \\
\midrule
Imaging (3) & Standardized format (NIfTI), per image metadata sidecar, consistent file naming \\
\midrule
Annotation (4) & Structured annotation directory, segmentation masks, per annotation metadata sidecar, machine readable label map \\
\midrule
Provenance (5) & Annotator identity, annotator credentials, annotation tool, annotation date, QC review documented \\
\midrule
Quality (2) & Inter annotator agreement, quality summary \\
\midrule
ML Readiness (2) & Documented splits, split rationale \\
\bottomrule
\end{tabular}
\end{table}

\subsection{Datasets Analyzed}

Four widely used public datasets were analyzed:

\begin{itemize}
    \item \textbf{LIDC-IDRI}~\citep{armato2011lung}: 1,018 chest CT
    scans with XML encoded nodule annotations from four radiologists.
    \item \textbf{BraTS}~\citep{brats2023}: Multi institutional
    brain MRI glioma segmentation challenge dataset with expert
    annotations.
    \item \textbf{CheXpert}~\citep{irvin2019chexpert}: 224,316 chest
    radiographs with NLP derived labels from radiology reports.
    \item \textbf{Medical Segmentation Decathlon
    (MSD)}~\citep{antonelli2022medical}: 2,633 training cases across 10
    segmentation tasks in multiple modalities.
\end{itemize}

\subsection{Results}

Table~\ref{tab:compliance} presents the compliance analysis
results.\footnote{Scores were assessed from publicly available dataset
documentation and file structures. Per dimension scoring criteria are
available at the project repository.}

\begin{table}[h]
\centering
\caption{Compliance analysis: four public datasets scored against 22
VIDS dimensions. Scores show satisfied/total per category. Partial
scores count as 0.5.}
\label{tab:compliance}
\small
\begin{tabular}{lccccc}
\toprule
\textbf{Category} & \textbf{LIDC-IDRI} & \textbf{BraTS} & \textbf{CheXpert} & \textbf{MSD} & \textbf{VIDS native} \\
\midrule
Structure (6)      & 1.5 & 2.0 & 1.5 & 1.5 & 6 \\
Imaging (3)        & 1.0 & 2.0 & 1.0 & 2.0 & 3 \\
Annotation (4)     & 1.5 & 2.0 & 1.0 & 2.0 & 4 \\
Provenance (5)     & 1.0 & 0.5 & 0.0 & 0.0 & 5 \\
Quality (2)        & 1.0 & 1.0 & 0.0 & 0.0 & 2 \\
ML Readiness (2)   & 0.0 & 1.0 & 1.0 & 1.0 & 2 \\
\midrule
\textbf{Total (22)} & \textbf{6.0} & \textbf{8.5} & \textbf{4.5} & \textbf{6.5} & \textbf{22} \\
\textbf{Percentage}  & \textbf{27\%} & \textbf{39\%} & \textbf{20\%} & \textbf{30\%} & \textbf{100\%} \\
\bottomrule
\end{tabular}
\end{table}

\subsection{Key Findings}

\textbf{Provenance is the largest gap.} Across all four datasets, the
provenance category averaged 0.4/5 (8\%). LIDC-IDRI documents its
annotation protocol in the original publication~\citep{armato2011lung}
but anonymizes individual reader identities and provides no
per annotation tool or date metadata in machine readable form. BraTS,
CheXpert, and MSD provide no per annotation provenance at all. It is
important to note that this is not a criticism of those datasets. Rather,
it reflects the absence of a standard that would have required such
documentation.

\textbf{Quality documentation is absent or external.} LIDC-IDRI and
BraTS have inter annotator agreement data that is computable from the
raw annotations or published in challenge papers, but neither provides
it as a structured file within the dataset itself. CheXpert and MSD
provide no quality documentation.

\textbf{Structure and imaging are partially addressed.} All four
datasets have some directory organization and imaging data in
standardized formats, but none use a machine readable dataset marker,
structured participant registry, or per image metadata sidecar.

\textbf{ML readiness varies.} MSD, BraTS, and CheXpert provide
predefined splits, though documentation of split rationale and leakage
prevention varies. LIDC-IDRI provides no predefined splits.

The average compliance across all four datasets is 29\%. The systematic
gaps, particularly in provenance (8\%) and quality documentation
(25\%), are precisely the information needed for regulatory
submissions, reproducibility, and buyer trust.

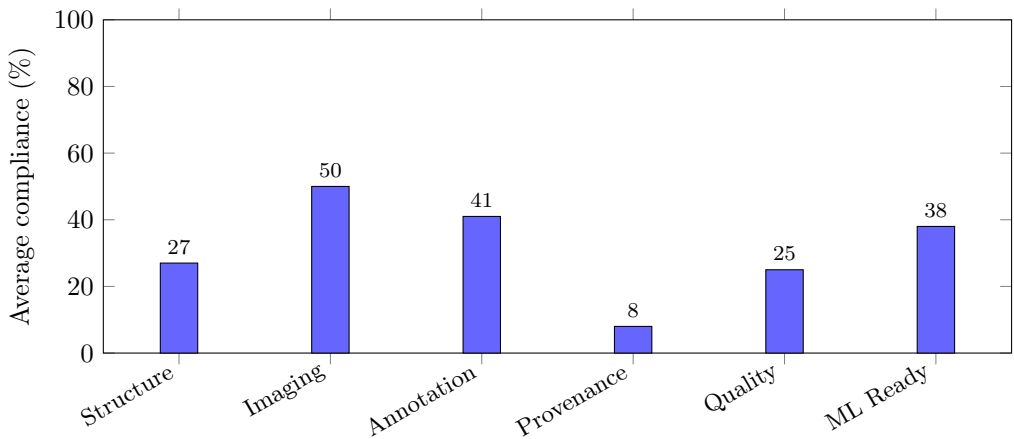
\begin{figure}[h]
\centering
\begin{tikzpicture}
\begin{axis}[
    ybar,
    width=0.85\textwidth,
    height=6cm,
    ylabel={Average compliance (\%)},
    symbolic x coords={Structure, Imaging, Annotation, Provenance, Quality, ML Ready},
    xtick=data,
    x tick label style={rotate=30, anchor=east, font=\small},
    ymin=0, ymax=100,
    bar width=14pt,
    nodes near coords,
    nodes near coords style={font=\footnotesize},
    every node near coord/.append style={anchor=south},
]
\addplot[fill=blue!60] coordinates {
    (Structure,27) (Imaging,50) (Annotation,41) (Provenance,8) (Quality,25) (ML Ready,38)
};
\end{axis}
\end{tikzpicture}
\caption{Average compliance across four public datasets by category.
Provenance and quality documentation represent the largest systematic
gaps.}
\label{fig:compliance_chart}
\end{figure}

Figure~\ref{fig:compliance_chart} visualizes the per category averages,
where each bar represents the mean score across all four datasets
expressed as a percentage of the category maximum.

\section{LIDC-Hybrid-100: A VIDS Compliant Reference Dataset}
\label{sec:case_study}

In order to demonstrate VIDS in practice, a subset of LIDC-IDRI was
converted into a fully VIDS compliant reference dataset: LIDC-Hybrid-100.

\subsection{Pipeline}

The conversion pipeline comprises eight steps executed sequentially:

\begin{itemize}
    \item[(i)] \textbf{Subject selection:} 1,010 LIDC-IDRI subjects were
    scanned via DICOM headers (one header per series, file count for
    slices). The subjects were filtered to CT modality with a minimum of
    100 slices, ranked by slice thickness (thinnest first), and the top
    100 were selected. It is noteworthy that selection was based solely on
    acquisition characteristics (slice thickness, slice count) and not
    on annotation quality, nodule count, or clinical outcomes.
    \item[(ii)] \textbf{DICOM to NIfTI conversion:} Selected subjects
    were converted using dcm2niix. Full DICOM metadata was captured
    (66 PHI tags stripped, all non PHI tags preserved).
    \item[(iii)] \textbf{VIDS structure:} The subject/session/modality
    directory tree was built. VIDS compliant imaging sidecars were
    generated from captured DICOM metadata.
    \item[(iv)] \textbf{Scaffolding:} Root metadata files were generated,
    viz., \texttt{dataset\_description.json}, \texttt{participants.json},
    \texttt{README.md}, and \texttt{CHANGES.md}.
    \item[(v)] \textbf{Annotation generation:} LIDC-IDRI XML annotations
    were parsed via pylidc. For each nodule cluster with $\geq$2 readers,
    per reader masks were extracted, consensus segmentation was computed
    ($\geq$50\% reader agreement), and pairwise Dice coefficients were
    calculated. Segmentation masks and provenance sidecars were written
    in a single pass.
    \item[(vi)] \textbf{Full profile documentation:} Quality summary,
    inter annotator agreement documentation, and ML splits (70/15/15
    train/val/test, subject level, seed~42) were generated. The profile
    was upgraded to Full.
    \item[(vii)] \textbf{Validation:} The 21 rule validator was executed.
    \item[(viii)] \textbf{Packaging:} A compressed archive with SHA-256
    checksum was created for Zenodo upload.
\end{itemize}

Consensus segmentation masks were generated from four radiologist
LIDC-IDRI XML annotations using majority vote ($\geq$50\% reader
agreement per voxel), with pairwise Dice coefficients computed from all
reader pair combinations for inter annotator agreement.

\subsection{Results}

Table~\ref{tab:lidc_results} summarizes the LIDC-Hybrid-100 dataset.

\begin{table}[h]
\centering
\caption{LIDC-Hybrid-100 dataset summary.}
\label{tab:lidc_results}
\small
\begin{tabular}{ll}
\toprule
\textbf{Metric} & \textbf{Value} \\
\midrule
Total subjects                & 100 \\
Subjects with annotations     & 89 \\
Subjects without annotations  & 11 (no qualifying nodules) \\
Total nodules                 & 224 \\
Nodules per subject (mean)    & 2.5 (range: 1--10) \\
\midrule
Mean pairwise Dice            & 0.7765 \\
Min pairwise Dice             & 0.3639 \\
Max pairwise Dice             & 0.9378 \\
\midrule
Quality: Excellent ($\geq$0.90)   & 3 subjects \\
Quality: Good ($\geq$0.85)        & 14 subjects \\
Quality: Acceptable ($\geq$0.75)  & 45 subjects \\
Quality: Poor ($<$0.75)           & 27 subjects \\
\midrule
Slice thickness range         & 0.60--1.25~mm \\
Manufacturers                 & Siemens (68), Philips (29), GE (3) \\
ML splits                     & Train: 70, Val: 15, Test: 15 \\
\midrule
Archive size                  & 12.5~GB (387 files) \\
VIDS validation               & 21/21 PASS (Full profile) \\
\bottomrule
\end{tabular}
\end{table}

The 11 subjects without annotations had only single reader nodule
annotations in the LIDC-IDRI source, preventing consensus mask
generation and inter annotator agreement computation ($\geq$2 readers
required). These subjects retain imaging files and participate in ML
splits.

\subsection{Provenance Characteristics}

The LIDC-IDRI source anonymized individual reader identities. Rather
than fabricating provenance, the VIDS sidecars document this
transparently:

\begin{itemize}
    \item \textbf{Annotator:} ``LIDC-IDRI Radiologist Panel,''
    board certified radiologists, 4 readers per scan, 7 institutions
    \item \textbf{Process:} LIDC annotation interface, date range
    2004--2010, consensus method and threshold documented
    \item \textbf{QC:} LIDC two phase blinded review, per subject mean
    Dice as confidence score
\end{itemize}

This brings to limelight a key VIDS design principle: provenance
documentation makes gaps \textit{visible}. The thin provenance (anonymous
readers, no per case dates) passes validation but is immediately apparent
to any buyer or reviewer examining the sidecar. Hence, VIDS makes quality
assessable, not guaranteed.

\subsection{Validation}

The complete validation output:

\begin{lstlisting}[language=bash,basicstyle=\ttfamily\footnotesize]
$ vids-validate LIDC-Hybrid-100/ --profile full
  S001-S006: PASS (100 subjects, all with sessions)
  I001-I004: PASS (100 imaging files, 100 sidecars)
  A001-A005: PASS (89 segmentation files, provenance complete)
  Q001-Q003: PASS (quality summary + agreement)
  M001-M002: PASS (splits.json)
  D001:      PASS (CHANGES.md present)
  VALIDATION PASSED (21/21 rules)
\end{lstlisting}

\section{Regulatory Context}
\label{sec:regulatory}

VIDS does not claim regulatory compliance. It provides auditable,
machine readable artifacts aligned with regulatory expectations for
data traceability and governance. The EU AI Act (Article~10) mandates
data governance, quality, representativeness, and bias control for
high risk AI systems. The IMDRF Good Machine Learning Practice (GMLP)
framework, authored by 10 international regulators, requires data
quality, provenance, and annotation protocols. The US FDA AI/ML SaMD
Action Plan emphasizes traceability and lifecycle documentation.

The 21 VIDS rules map to these requirements:
\texttt{dataset\_description.json} (S002) provides provenance and
traceability; \texttt{participants.json} (S003) enables demographic bias
analysis; annotation sidecars (A005) document the annotation chain of
custody; quality documentation (Q001--Q003) provides inter annotator
agreement evidence; and \texttt{splits.json} (M002) prevents data
leakage.

Therefore, a VIDS validation report does not certify regulatory
compliance, but it provides structured, machine readable evidence that
can be attached to a regulatory submission as documentation of data
governance practices.

Furthermore, VIDS compliance makes it transparent whether a dataset has
received ethical approval. The \texttt{Compliance} object in
\texttt{dataset\_description.json} documents IRB approval status and
de identification methods. A dataset without ethical approval
documentation passes structural validation, but the absence is
immediately visible to any reviewer or regulatory authority examining
the metadata.

\section{Discussion}
\label{sec:discussion}

\subsection{Scope and Limitations}

VIDS v1.0 does not cover 4D temporal sequences (e.g., cardiac cine MRI),
non imaging clinical data (EHR, genomics), or real time annotation. NIfTI
is the current reference format; DICOM as a native storage format is
under consideration for v2.0.

The validator enforces structural compliance by design. Quality
assessment, viz., whether annotations are accurate, demographics
representative, or a dataset suitable for a particular clinical task,
is intentionally outside the validation scope. VIDS makes quality
assessable by requiring structured documentation; quality evaluation
tools built on VIDS metadata are a natural downstream application.

\subsection{Adoption Model}

VIDS follows the model of successful open standards: the specification
is CC~BY~4.0, tools are Apache~2.0, and governance is designed to
transition from single organization stewardship to consortium governance
as external adoption grows. The two profile design (POC and Full) lowers
the adoption barrier, and teams can start with 15 rules and graduate
to 21 as their processes mature.

\subsection{Future Work}

Two directions are planned. \textit{Ecosystem extensions}: a
\texttt{vids-clinical} package for clinical metadata linkage, a
synthetic data disclosure field (\texttt{DataOrigin}) in v1.1, and
framework native data loaders for MONAI and nnU-Net.
\textit{Evaluation layer}: quality scoring tools that operate on VIDS
metadata to compute dataset level quality grades from the structured
provenance and agreement data that VIDS already requires.

\section{Conclusion}
\label{sec:conclusion}

This paper presents VIDS, a standard that provides machine enforceable
validation across dataset structure, annotation provenance, quality
documentation, and ML readiness in a single framework. The compliance
analysis of four major public datasets reveals systematic gaps averaging
71\% across 22 dimensions, with provenance and quality documentation as
the weakest areas.

LIDC-Hybrid-100 demonstrates that existing datasets can be restructured
into VIDS compliant format with complete provenance documentation,
achieving 21/21 on the Full compliance profile. The dataset, validator,
and specification are freely available.

\paragraph{Availability.}~\\
Specification: \url{https://vidsstandard.org} \\
Validator: \texttt{pip install vids-validator} \\
Source: \url{https://github.com/vids-standard/vids-standard} \\
Dataset: \url{https://doi.org/10.5281/zenodo.19582717}

\bibliographystyle{unsrt}
\bibliography{references}

\end{document}